\documentclass[prb,twocolumn,superscriptaddress,floatfix,showkeys, letterpaper]{revtex4}
\usepackage{graphicx,amssymb,bm,natbib,amsfonts,amsmath}
\usepackage{hyperref}

\begin{document}

\title{Low energy excitations in graphite: The role of dimensionality and lattice defects}

\author{S.Y. Zhou}
\affiliation{Department of Physics, University of California,
Berkeley, CA 94720, USA}
\affiliation{Materials Sciences Division,
Lawrence Berkeley National Laboratory, Berkeley, CA 94720, USA}
\author{G.-H. Gweon}
\affiliation{Department of Physics, University of California,
Berkeley, CA 94720, USA}
\author{A. Lanzara}
\affiliation{Department of Physics, University of California,
Berkeley, CA 94720, USA}
\affiliation{Materials Sciences Division,
Lawrence Berkeley National Laboratory, Berkeley, CA 94720, USA}

\date{\today}

\begin{abstract}
In this paper, we present a high resolution angle resolved photoemission spectroscopy (ARPES) study of the electronic properties of graphite.  We found that the nature of the low energy excitations in graphite is particularly sensitive to interlayer coupling as well as lattice disorder.  As a consequence of the interlayer coupling, we observed for the first time the splitting of the $\pi$ bands by $\approx$0.7 eV near the Brillouin zone corner K.  At low binding energy, we observed signatures of massless Dirac fermions with linear dispersion (as in the case of graphene), coexisting with quasiparticles characterized by parabolic dispersion and  finite effective mass.  We also report the first ARPES signatures of electron-phonon interaction in graphite: a kink in the dispersion and a sudden increase in the scattering rate.  Moreover, the lattice disorder strongly affects the low energy excitations, giving rise to new localized states near the Fermi level.  These results provide new insights on the unusual nature of the electronic and transport properties of graphite.  
\end{abstract}

\keywords{graphite, graphene, Dirac fermions, low energy excitations, disorder, electron-phonon interaction, localized states}

\maketitle
Carbon, one of the most basic elements in nature, still gives a lot of surprises.  It is found in many different forms (``allotropes'') -- from zero dimensional fullerene, one dimensional carbon nanotubes, two dimensional graphene and graphite, to three dimensional diamond -- and the properties of the various carbon allotropes can vary widely \cite{Carbon}. For example, diamond is the hardest material, while graphite is one of the softest; diamond is transparent to the visible spectrum, while graphite is opaque; diamond is an electrical insulator while graphite is a conductor. What is interesting is that all these different properties originate from the same carbon atoms, simply with different arrangements of the atomic structure.

One of the prime examples for the unique properties of carbon is graphite.  Although graphite is one of the most widely studied materials, it continues to provide scientists with new and interesting challenges as recently testified by a wide range of novel discoveries in the field, such as novel quantum Hall effect \cite{Kopelevich}, room temperature ferromagnetism \cite{KopelevichPRB}, metal-insulator-like transition \cite{Kopelevich}, and superconductivity \cite{GIC}.  Also, the recent discovery that the electronic properties of thin graphite samples can be modified by externally applied voltage, in such a way that these systems can be switched from n-type to p-type carriers \cite{NovoselovSci, ZhangPRL}, has raised a great potential for carbon-based nanoelectronics.  The secret to all of these fascinating phenomena lies in the unique nature of the electronic properties of graphite, where the low energy excitations resemble massless Dirac fermions.  Although several transport experiments have shown results in agreement with the existence of Dirac fermions in graphite \cite{Kopelevich2}, so far direct experimental evidence has been limited. 

In this paper, we report an angle resolved photoemission spectroscopy (ARPES) study of the low energy excitations in graphite and we report signatures for the coexistence of massless Dirac fermions and quasiparticles with finite effective mass.  We investigate the effect of interlayer coupling and lattice disorder, which provides new insights on the nature of the electronic and tranport properties of graphite.

\section*{EXPERIMENTAL TECHNIQUE}

ARPES is one of the most widely used techniques to study the electronic structure of solids and is the only tool that can directly map the energy-momentum phase space of electrons.  This is based on the principle of the photoelectric effect, explained by A. Einstein in 1905.  By shining a beam of monochromatic light on the sample, electrons (``photoelectrons'') are emitted by the photoelectric effect (see schematic of a typical ARPES experiment in Figures 1(a,b)) and can be probed as a function of their kinetic energy for each given emission angle, which uniquely determines the momentum of the photoelectrons.  
Therefore, given the conservation laws for energy and momentum, the kinetic energy and in-plane momentum of the photoelectrons can be related unambiguously to the binding energy and the crystal momentum of electrons in a solid, providing unique information on the direction, the speed and the scattering processes of valence electrons.

For a two dimensional solid, where the electronic structure is anisotropic and there is no dispersion along the z axis (sample normal direction), the electronic dispersion is completely determined by the in-plane momentum k$_\parallel$.  On the other hand, for a three dimensional sample, the full momentum information including the out-of-plane momentum k$_z$ is important, since the electronic structure also depends on k$_z$.  Note that in the photoemission geometry, k$_\parallel$ is a good quantum number while k$_z$ is not.  Although this makes the interpretation of ARPES data for a three dimensional sample more difficult, a simple modelling of the photoemission process involving free-electron approximation of the final state photoelectrons has been successfully applied in ARPES \cite{Himpsel,  Hufner} to obtain full momentum information, including the k$_z$ value.  In particular, at normal emission (k$_\parallel$=0), k$_z$ is related to the kinetic energy E$_k$ by k$_z$=$\sqrt{2m(E_k+V_{in})/\hbar^2}$, where the inner potential V$_{in}$ is a parameter that can be determined from the symmetry of the measured dispersions \cite{Himpsel, Hufner}. 

\begin{figure}
\includegraphics[width=6.8 cm]{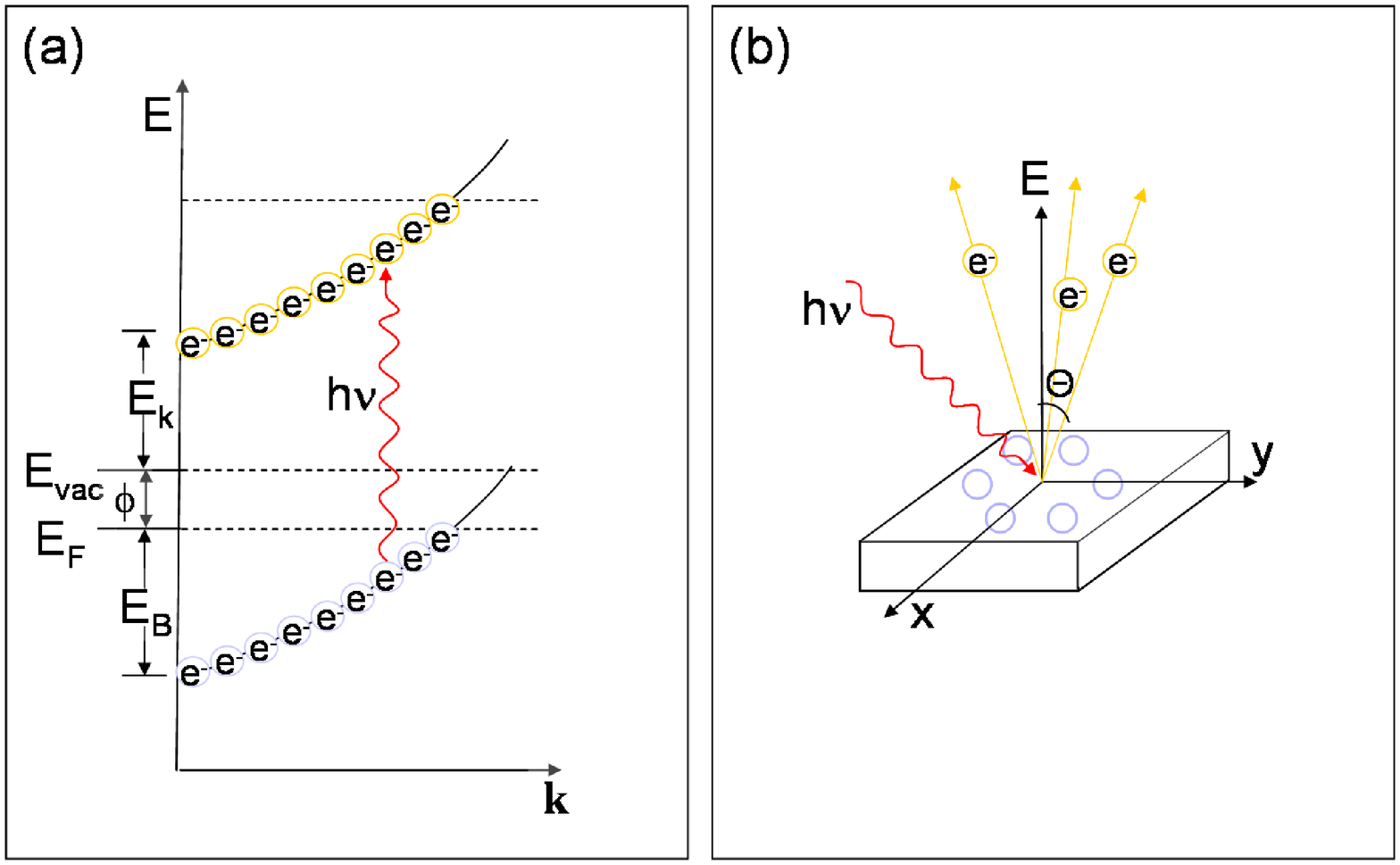}
\label{fig.1}
\caption {(a) Representation of a photoemission process.  Electrons with binding energy E$_B$ are excited to the vacuum and their kinetic energy E$_k$ is measured.  Here E$_F$ is the Fermi energy; E$_{vac}$ is the energy of the vacuum level and $\phi$ is the work function. (b) Typical geometry of an ARPES experiment.  Photoelectrons are measured as a function of E$_k$ and angle $\Theta$, which can be converted to binding energy E$_B$ and momentum k for electrons inside the solid.}
\end{figure}

Under the sudden approximation, ARPES measures the single particle spectral function A(k, $\omega$), A(k, $\omega$)=(-1/$\pi$)ImG(k, $\omega$), and therefore contains unique information on the self energy $\Sigma$(k, $\omega$) of solids. In the non-interacting case, A(k, $\omega$) is simply a delta function $\delta$(k, $\omega$).  From an experimental point of view,  A(k, $\omega$) is directly related to the intensity measured in an ARPES experiment by I(k, $\omega$) = A(k, $\omega$)f($\omega$)$\left|M(k, \omega)\right|^2$, where f($\omega$) is the Fermi function, reflecting the fact that photoemission can probe only the occupied states, and M(k, $\omega$), known as matrix element, depends on the electron momentum and on the energy and polarization of the incoming photon.  Therefore, by measuring the intensity of the photoemitted electrons as a function of their kinetic energy and emission angles, one can probe the electronic structure of the material \cite{Hufner}. 

A typical two dimensional ARPES spectrum is shown in Figure 2.  The color scale represents the intensity of the emitted photoelectrons, which is plotted as a function of their binding energy and in-plane momentum in panel a.  A traditional and powerful method of analyzing ARPES data involves energy distribution curves (EDCs), energy scans at constant momentum (panel b).  In principle, an EDC analysis can provide direct information on the total self energy $\Sigma$(k, $\omega$) of the system in two dimensions.  On the other hand, however, the EDC lineshape is complicated due to the presence of elastic and inelastic photoelectron scatterings in ARPES and the Fermi function, f($\omega$).  An alternative way to the EDC analysis is to analyze the momentum distribution curves (MDCs), momentum scans at constant energy (panel c).  One of the advantages of the MDC analysis is that the lineshape can be well approximated by a Lorentzian function.  However, an important caveat of this analysis is that it is often based on the unchecked assumptions of linear dispersion and momentum-independent self energy. 

\begin{figure}
\includegraphics[width=8.4 cm]{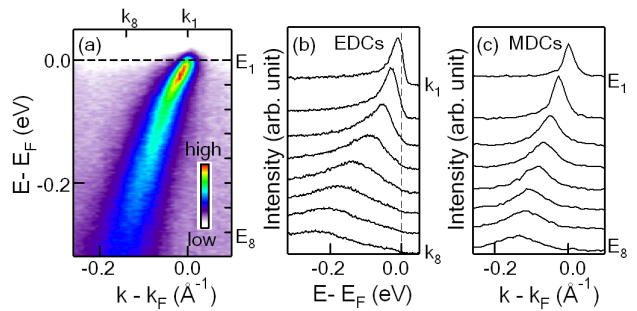}
\label{fig.2}
\caption {(a) Typical ARPES intensity map measured on a high temperature superconductor Bi2212, as a function of energy and momentum.  Different colors represent the intensity of photoelectrons.  The color scale for the intensity maps throughout this paper except Figures 5 and 11 is shown in the inset of panel a. (b) Energy distribution curves (EDCs) for momentum values from k$_0$ to k$_8$, indicated by vertical tick marks on the top of panel a.  (c) Momentum distribution curves (MDCs) taken at energies from E$_1$ and E$_8$, indicated by horizontal tick marks on the right of panel a.}
\end{figure}

In this review, we present high resolution ARPES data on graphite taken under different experimental conditions (polarization and photon energy) with main focus on the high symmetry directions.  Given its quasi-two dimensional crystallographic structure and the weak Van der Waals bonding between graphene planes, graphite is an ideal system to be studied by photoemission spectroscopy.  High resolution ARPES data have been collected at beam lines 10.0.1 and 12.0.1 of the Advanced Light Source of the Lawrence Berkeley National Laboratory.  The energy resolution is from 15 meV to 40 meV, and the angular resolution is $\le$ 0.3 degree.  The samples were cleaved {\em in situ} in ultra-high vacuum better than 6.0$\times$10$^{-11}$ Torr and measured at a temperature of 25 K\@.  The Fermi energy was calibrated by measuring the Fermi edge of Au.

The ARPES data presented in this paper have been recorded on highly oriented pyrolytic graphite (HOPG).  We note that, although in order to perform momentum resolved measurements, it is most desirable to have a properly aligned single crystal, we have recently shown that ARPES study can be performed also on an azimuthally disordered sample \cite{ZhouPRB}.  Almost paradoxically, the complete angular averaging gives a total dominance of the dispersions along the two high symmetry directions over dispersions in all other directions, and thus surprisingly sharp dispersions are observed in the ARPES data obtained from HOPG samples.  More mathematically, this has been explained in terms of the van-Hove singularities in the azimuthal density of states along the high symmetry directions \cite {ZhouPRB}, $\Gamma$-M-$\Gamma$$^\prime$ (A-L-A$^\prime$) and $\Gamma$-K-M$^\prime$ (A-H-L$^\prime$).  For the main interest of this paper, i.e.~the low energy electronic structure within 1.5 eV below E$_F$, the dispersions are dominated only by those along $\Gamma$-K-M$^\prime$ (A-H-L$^\prime$), since the dispersions along the other direction $\Gamma$-M-$\Gamma$$^\prime$ (A-L-A$^\prime$) lie at much higher binding energy (larger than 2.4 eV) \cite{NoteBS}. 
 
\section*{BAND STRUCTURE AND LOW ENERGY EXCITATIONS IN GRAPHITE}
Graphite can be considered as a quasi-2D system from both a structural and an electronic point of view.  From a structural point of view, the basic unit of graphite, graphene, is formed by carbon atoms arranged in a honeycomb lattice.  Graphite is formed by infinite layers of graphene stacked in an ABAB sequence.  In graphite, the in-plane C-C bonds are strong due to the small in-plane lattice spacing of 1.42 $\AA$, while the out-of-plane coupling is much weaker due to the large interlayer spacing of 3.37 $\AA$\@.  Thus the interlayer interaction in graphite is dominated by weak van der Waals force.  These structural properties impart quasi-two dimensionality to the electronic structure of graphite.
 
\begin{figure}
\includegraphics[width=8.4 cm]{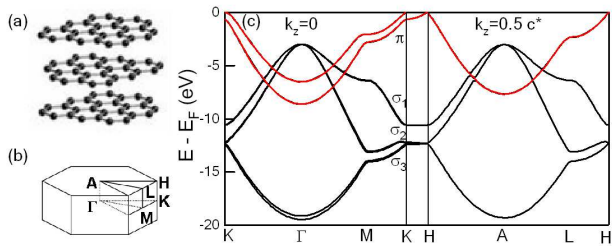}
\label{fig.3}
\caption {(a) Schematic drawing of the crystallographic structure of graphite. (b) Three dimensional Brillouin zone (BZ) of graphite.  (c) LDA dispersion for $\pi$ (red) and $\sigma$ (black) bands along the high symmetry directions for graphite.  The $\pi$ bands at low energy are the main focus of this paper and are highlighted by red color.}
\end{figure}

The electronic properties of graphite are highly anisotropic \cite{McClure, Tatar, Louie, Charlier, AntonioML}.  In graphite, the strong in-plane $\sigma$ bonds formed by 2s, 2p$_x$ and 2p$_y$ orbitals result in 3 $\sigma$ bands at high binding energy.  The out-of-plane $\pi$ bonding formed by 2p$_z$ orbitals is much weaker, and the $\pi$ bands are at much lower binding energy.  Among all these bands, only the $\pi$ bands cross the Fermi energy E$_F$ at the corners of the hexagonal Brillouin zone (BZ).  These low energy $\pi$ bands play the most important role in determining the electronic and transport properties of graphite and will be the focus of this paper.  Figure 3(c) shows the theoretical band structure of graphite.  We note that different from a perfect two dimensional graphene, the weak interlayer interaction in graphite results in a splitting of the $\pi$ bands in the k$_z$=0 plane (K-$\Gamma$-M-K).  While in the plane of k$_z$=0.5 c$^*$ (H-L-A-H), where c$^*$ is the reciprocal lattice constant along the stacking direction, the $\pi$ bands are degenerate, and the electronic structure in this k$_z$ plane strongly resembles that of graphene.  Despite the splitting of the $\pi$ bands, the overall dispersion of the $\pi$ band still bears a strong similarity to that of graphene.

\begin{figure}
\includegraphics [width=7.5 cm] {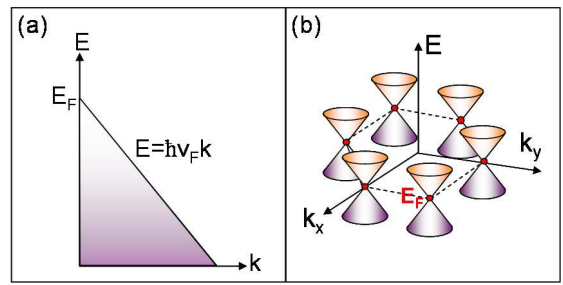}
\label{fig.4}
\caption{(a) Linear dispersion near E$_F$, which is expected for graphene.  (b) Point-like Fermi surface and cone-like dispersion near BZ corner.}
\end{figure}

The unique low energy electronic structure of graphene and graphite \cite{Fradkin} is of special interest and believed to be responsible for various exotic properties observed \cite{Zhang, Novoselov, Kopelevich, Xu}.   For most solids, the physics is successfully described in terms of the nonrelativistic Schr\"odinger equation.  The low energy excitations are quasiparticles with a finite effective mass, characterized by a finite Fermi surface and finite density of state at E$_F$.  On the other hand, graphene and graphite are semi-metals and the $\pi$ bands are believed to disperse linearly and touch only at one point near E$_F$ (Figure 4(a)), forming cone-like dispersions near each BZ corner (Figure 4(b)).  As a consequence, the Fermi surface is characterized by ``Fermi points'' (red dots in Figure 4(b)) located at the six corners of the hexagonal BZ, rather than a Fermi surface.  In this case, the low energy excitations resemble relativistic Dirac fermions and are essentially described by the relativistic Dirac equation with the effective speed of light replaced by the Fermi velocity.  The density of states is vanishing at the crossing point (known as Dirac point), leading to many peculiar phenomena in these materials,  e.g.~novel quantum Hall effect \cite{Zhang, Novoselov, Kopelevich}, magnetic field driven metal-insulator-like transition \cite{Kopelevich}, unconventional lifetime \cite{Xu, Gonzalez, Catalin}, and spin-Hall effect in graphene \cite{SpinHall}. 

\section*{EXPERIMENTAL RESULTS}
\section*{I. BAND STRUCTURE AND INTERLAYER COUPLING}

\begin{figure}
\includegraphics[width=8.8cm]{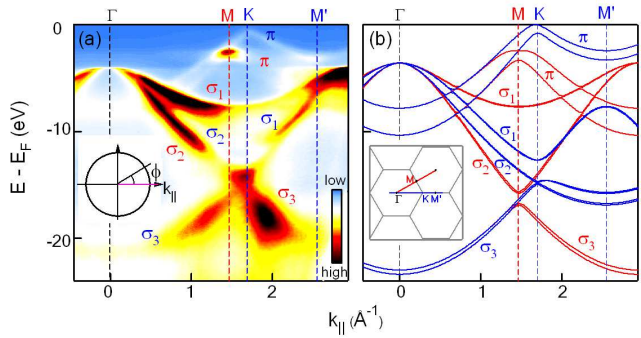}
\label{fig.5}
\caption {(a) ARPES intensity map taken at 60 eV photon energy (k$_z$ $\approx$ 0.28 c$^*$) on HOPG along the gray line shown in the inset.  The color scale for this figure and Figure 11 is shown in the inset.  (b) LDA dispersions along two high symmetry directions $\Gamma$KM$^\prime$ (blue lines) and $\Gamma$M (red lines) stretched by 20$\%$ in energy.}
\end{figure}

Figure 5(a) shows the measured ARPES intensity map as a function of binding energy and in-plane momentum $k_\parallel$ over a wide energy range (from E$_F$ to -22 eV).  In this color scale, red is maximum intensity and blue is zero intensity.  Despite the azimuthal disorder nature of the sample, sharp dispersions from two high symmetry directions $\Gamma$KM$^\prime$ and $\Gamma$M directions can still be clearly resolved.  The overall dispersion is in good agreement with the LDA band structure stretched by 20$\%$ in energy (shown in panel b for comparison), and in agreement with previous measurements \cite{Marchand, NarrowLines, Daimon}.  The contributions from different orbitals $\pi$, $\sigma$ can be clearly distinguished.  We now focus on the low energy $\pi$ bands between E$_F$ and -11 eV, the only bands that cross E$_F$.

\begin{figure}
\includegraphics[width=7.5 cm]{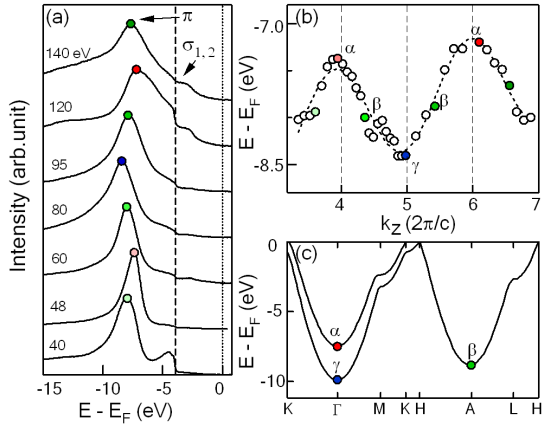}
\label{fig.6}
\caption {(a) Angle-integrated intensity curves taken near normal emission measured at different photon energies from 40 to 140 eV.  Filled circles mark the peak positions of the $\pi$ bands.  (b) Extracted peak positions from the angle-integrated intensity curves as a function of k$_z$.  The dotted line is the guide to the periodicity of the dispersion.  From the symmetry of the final states detected, the inner potential is determined. (c) LDA band structure of the $\pi$ bands at k$_z$=0 and k$_z$=0.5 c$^*$.  The energies are stretched by 20$\%$.}
\end{figure}

In order to investigate the details of the low energy electronic structure such as the effect of interlayer coupling, it is important to obtain the full momentum information including the out-of-plane momentum k$_z$.  To extract the k$_z$ value, we first determine the inner potential V$_{in}$ from the symmetry of the measured dispersion along $\Gamma$A (k$_\parallel$=0).  Figure 6(a) shows a few examples of angle-integrated EDCs over a wide photon energy range from 34 to 155 eV, from which the dispersion is extracted.  There are two features in panel a, a main peak at higher binding energy associated with the $\pi$ bands, and a hump at lower binding energy associated with the $\sigma$ bands.  The main peak from $\pi$ bands shows oscillation between -7.2 and -8.4 eV as a function of photon energy, or equivalently k$_z$.  The extracted dispersion (panel b) from the peak positions can be described as a periodic oscillation riding on a linear slope.  The linear slope can be explained by angle integration over 8 degrees \cite {fnkzdis} due to the angle average mode used for this data set alone among the data shown in this paper.  The periodic behavior and the symmetry of the dispersion enable us to determine the inner potential (17$\pm$1 eV) and thus extract the k$_z$ values.  The periodic dispersion bears a strong similarity with previously reported data in the literature \cite {Pescia, Law}, but covers more than twice the k$_z$ range.  We note that, as previously reported \cite{Pescia, Law}, the dispersion shows a periodicity of 2 c$^*$ rather than c$^*$.  This doubling periodicity has been observed \cite{Law} and can be understood by combining the k$_z$ dispersion of $\pi$ bands shown in panel c and the symmetry of the final states detected using the dipole selection rules \cite{BZselection}.  More specifically, due to the nonsymmorphic group in graphite, the final states detected changes in a repeated sequence of  $\alpha-\beta-\gamma-\beta-\alpha$.  We note that $\alpha$ and $\gamma$ states are resulted from the splitting of the $\pi$ bands, and thus the measured dispersion at the zone center is a reflection of the interlayer coupling.

\begin{figure}
\includegraphics [width=6.8 cm] {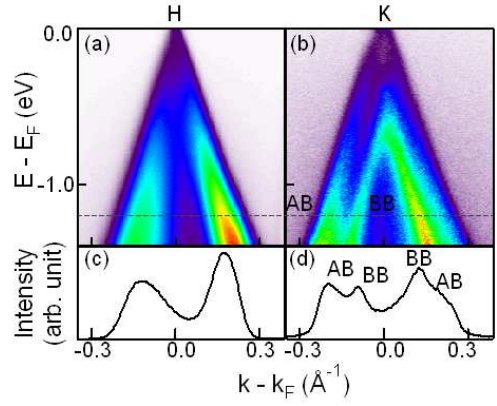}
\label{fig.7}
\caption{(a-b) ARPES intensity map measured near the BZ corners at photon energies of 43 (k$_z$ $\approx$ 0.35 c$^*$) and 55 eV (k$_z$ $\approx$ 0.10 c$^*$) respectively.  AB and BB label the antibonding and bonding $\pi$ bands.  (c-d) MDCs at -1.2 eV for data shown in panels a and b respectively.}
\end{figure}

We now focus on the effect of the interlayer coupling near the BZ corners.  Figure 7 shows the $\pi$ bands as a function of k$_\parallel$ near the BZ corner for two different k$_z$ values, 0 and 0.5 c$^*$.  Panel a shows an ARPES intensity map of the $\pi$ bands taken near the H point (k$_z$=0.5 c$^*$).  The data show a $\Lambda$-shaped dispersion crossing E$_F$ near the apex (H point), similar to the dispersion of the Dirac fermions.  The Fermi velocity v$_F$ estimated from the slope of the dispersion dE/dk=6 eV$\cdot\AA$ is v$_F$= 0.91$\pm$0.15 m$\cdot$s$^{-1}$, similar to reported values on graphene by transport measurements \cite{Zhang, Novoselov}.  Panel b shows an ARPES intensity map taken near the K point (k$_z$=0).  In this case, one can clearly distinguish two $\Lambda$-shaped dispersions, one similar to that in panel a, and the other located at a higher binding energy.  The intensity variation in the bonding and antibonding bands is attributed to the photoemission dipole matrix element \cite {BZselection}.  The splitting of the $\pi$ bands can be confirmed in the MDC shown in panel d, where two peaks each from the bonding and antibonding $\pi$ bands are clearly observed.  From the MDC dispersions shown and the peak positions in the EDC at the K point (not shown), this splitting is estimated to be $\approx$ 0.7 eV\@.  This is the first clear demonstration of the splitting of the $\pi$ bands near E$_F$, while we note that some data in the literature \cite {Law, PiBandSplit, NarrowLines} may now be seen as suggestive of this splitting.  The splitting of $\pi$ bands near the K point but not near the H point is in agreement with band structure calculation \cite{McClure, Tatar, Louie, Charlier, AntonioML}, which also confirms the validity of extracting the k$_z$ values.  Furthermore, the linear $\Lambda$-shaped dispersions shown in panel a, strongly resembling those of Dirac fermions, are signatures of Dirac fermions.

\section*{II. LOW ENERGY EXCITATIONS AND DIRAC FERMIONS}

We now focus on the nature of the low energy excitations in graphite.  These are of fundamental importance to understand transport properties as well as various exotic behaviors observed in graphene and graphite \cite{Zhang, Novoselov, Kopelevich, Xu}.  Recently, tranport studies of graphene have shown that the low energy excitations behave like Dirac fermions.  One interesting finding is that the types of charge carriers can be tuned from electron to hole by applying an electric (magnetic field), due to the vanishing density of states near E$_F$ \cite{NovoselovSci, ZhangPRL}.  In graphite, transport measurements have suggested two main types of excitatioins (massless Dirac fermions and quasiparticles with finite effective mass) \cite{Kopelevich2}, while direct investigation on the properties of these low energy excitations has been missing so far.

\begin{figure}
\includegraphics[width=7.5 cm]{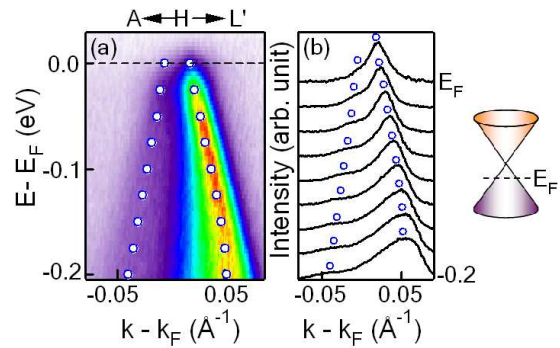}
\label{fig.8}
\caption{(a) ARPES intensity map taken at 20 eV photon energy  (k$_z$ $\approx$ 0.46 c$^*$) near H. The blue open circles are the dispersions extracted from MDCs.  (b) MDCs at energies from -0.2 eV to E$_F$ measured from data shown in panel a.  The schematic drawing on the right hand side of panel b summarizes the dispersions near H.} 
\end{figure}

Figure 8 shows high resolution ARPES data measured along AHL$^\prime$ direction.  By following the maximum intensity in panel a, it is clear that the dispersion shows a linear behavior.  In panel b, we show the raw MDCs at different binding energies.  In all the MDCs, one can clearly distinguish two peaks, a main peak and a weaker one.  The dispersion extracted from MDC peak positions (open circles in panels a,b) shows a linear behavior, strongly resembling the behavior of Dirac fermions.  The Fermi velocity, or the effective speed of light, is determined to be 0.8$\pm$0.2$\times$10$^{-6}$ m$\cdot$s$^{-1}$, also similar to values reported from band structure calculation \cite{GICDresselhaus} and transport measurements \cite{Novoselov}.  By extrapolating the dispersion, the crossing point of these two linear bands (known as the Dirac point) lies above E$_F$, suggesting that the low energy excitations near H point are holes, which is in agreement with transport measurements \cite{Kopelevich2, DresselhausPRL}.  Moreover, with the advantage of ARPES to directly probe the momentum space, the location of the hole pocket as well as the Dirac fermions can be determined directly to be near the H point.

\begin{figure}
\includegraphics[width=8.2 cm]{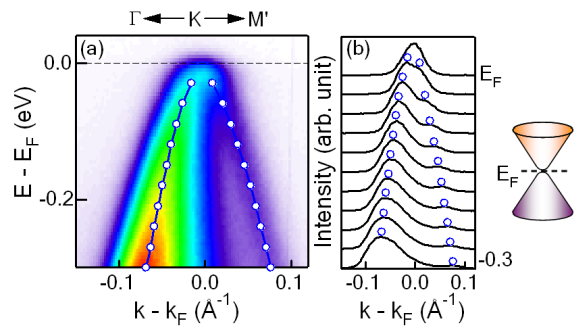}
\label{fig.9}
\caption{(a) ARPES intensity map near the zone corner K measured at 50 eV photon energy (k$_z$ $\approx$ 0.08 c$^*$).  Due to the different matrix elements caused by different photon energies and experimental geometries, the intensity in this case is strongly enhanced along $\Gamma$K direction.  Open circles are the dispersions extracted from MDCs.  The inset shows a schematic drawing of the dispersion near K.  (b) MDCs from -0.3 eV to E$_F$. The schematic drawing on the right hand side of panel b summarizes the dispersions near K.} 
\end{figure}

\begin{figure*}
\includegraphics[width=12.0 cm]{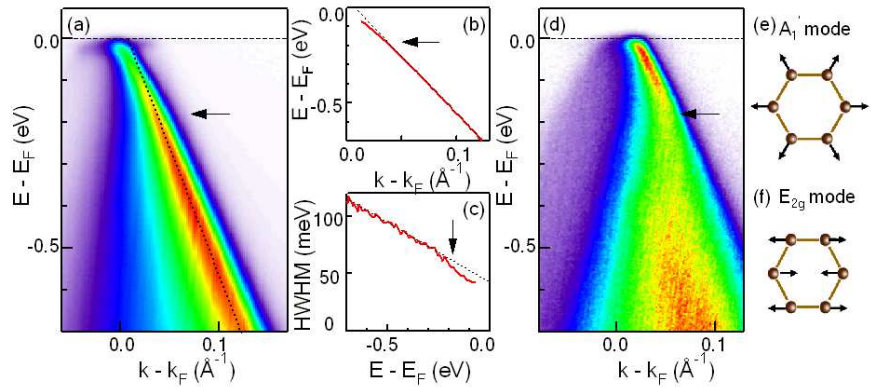}
\label{fig.10}
\caption{(a) ARPES intensity map taken at 26 eV photon energy (k$_z$ $\approx$ 0.13 c$^*$). (b) EDC Dispersion and (c) EDC half width half maximum extracted from the data, showing that there is an ARPES kink at $\approx$ 180 meV. (d) APRES intensity map taken at 20 eV photon energy (k$_z$=0.46 c$^*$). (e,f) A$_1^\prime$ and E$_{2g}$ phonon modes, which might be involved in the kink observed in panels a and c.} 
\end{figure*}

To investigate how the low energy excitations evolve as a function of k$_z$, we show in Figure 9 the intensity map near the zone corner K (k$_z$=0), along $\Gamma$KM$^\prime$ direction.  In contrast to the linear behavior observed in Figure 8 near the H point, the dispersion shows clearly a parabolic behavior.  The parabolic dispersion can also be followed by tracking the peak positions in the MDCs (panel b).  The parabolic behavior observed here at k$_z$=0 clearly suggests that the quasiparticles have a finite effective mass.  Further analysis to extract the effective mass is in progress \cite{ZhouSubmitted}.

To conclude this part, the data shown in Figures 8-9 suggest that the low energy excitations in graphite are: massless Dirac fermions near the H point; quasiparticles with finite effective mass near the K point \cite{ZhouSubmitted}.

\section*{III. ELECTRON-PHONON INTERACTION}

Electron-phonon interaction is important in understanding the properties of not only graphite, but also carbon nanotubes which can be considered as rolled-up graphene sheets.  First, electron-phonon interaction is important in resolving the long debate on the nature of Raman D and G peaks observed in carbons \cite{Ferrari}.  Moreover, electron-phonon interaction might be responsible for the breakdown of the ballistic transport expected for quasi-one-dimensional carbon nanotubes \cite{Yao, Javey}.  Recently, electron-phonon interaction has been reported in the phonon dispersion of graphite, appearing as two Kohn anomalies for the E$_{2g}$ mode at $\Gamma$ and the A$_1^\prime$ mode at the K point \cite{PhononDis, KohnAno}.  On the other hand, evidence of electron-phonon interaction on the electronic structure has been missing so far.  To study the effect of electron-phonon interaction in the electronic structure, APRES can provide direct information, as has been successfully demonstrated in metals and high temperature superconductors \cite{Valla, Lanzara}.  Here we report the first APRES signature of electron-phonon interaction in graphite: a kink in both the dispersion and the scattering rate.

Figure 10(a) shows an ARPES intensity map near the K point.  The dispersion, while showing a general linear-like behavior, shows a non-trivial structure at $\approx$ 180 meV. This observation is made clear in panel b where the ARPES dispersion extracted from EDC peak positions is shown. As shown by the dotted line, the dispersion deviates from a linear behavior at energy $\approx$ -180 meV.  This type of dispersion change is typical of an ARPES kink that arises from the interaction of electron with a collective mode, but it also could be the result of a genuinely non-linear one electron band dispersion as discussed above. In order to distinguish the two possibilities, it is essential to examine the ARPES line width, which reflects the inverse lifetime, or the scattering rate. If the energy scale is caused by an ARPES kink, then it is expected that the line width shows a sudden increase at that energy scale, consistent with an increased incoherence at high energy, while in the one-electron band scenario, no such abrupt increase is expected. In panel c, we show the EDC width (half width at half maximum on the E$_F$ side) as a function of energy, clearly demonstrating that the ARPES kink scenario is more appropriate.  Figure 10(d) shows an ARPES intensity map measured near the H point.  Near E$_F$, the dispersion is linear as discussed in Figure 8, while a stronger kink in the dispersion and a broadening of the peak above the kink energy is clearly observed, suggesting that a stronger kink is present near the H point.

Now that we have shown that there is a clear ARPES kink signature in graphite, a natural question follows as to its origin. We find that electron-lattice coupling is a probable source of the kink observed, since the kink energy is similar to the energies of the phonons which are responsible for the Kohn anomalies in the phonon dispersion \cite{KohnAno}, $\approx$ 160 meV for A$_1^\prime$ mode (panel e) and $\approx$ 200 meV for E$_{2g}$ mode (panel f).  Further study to identify the boson modes responsible for this kink and investigate the momentum dependence of this ARPES kink is in progress.

\section*{VI. LOCALIZED STATES IN GRAPHITE}

Within the graphene plane or each stacked graphene layer in graphite, a varying number of imperfections can be found, such as vacancies when lattice sites are unfilled indicating a missing atom within a basal plane; stacking faults when the ABAB sequence of the graphene planes is no longer maintained; and zigzag edge occurring near unfilled lattice sites \cite{Castroneto, DefectsGraphene}.  Because of the unique electronic structure of Dirac fermions and the negligible density of states near the Dirac point, graphite is extremely sensitive to topological defects which can strongly modify the electronic structure and the scattering process of quasiparticles, thus resulting in important changes in the electronic and transport properties.  For example, it has been predicted \cite{Castroneto} that extended defects as lattice dislocation can lead to self-doping effects and presence of localized states at the Fermi energy.  Self-doping effect results in electron or hole pocket at E$_F$ rather than a single point, as predicted for an ideal graphene system.  Zigzag edge on the other hand induces localized states near the Fermi energy, resulting in a peak in the local density of states near the Fermi energy \cite{Dresselhaus, STMedges}.  Finally the presence of disorder strongly modifies the scattering process, resulting in a minimum of the scattering rate at finite binding energy \cite{Castroneto} instead of E$_F$ as what is expected for Fermi liquid.

\begin{figure}
\includegraphics[width=7 cm]{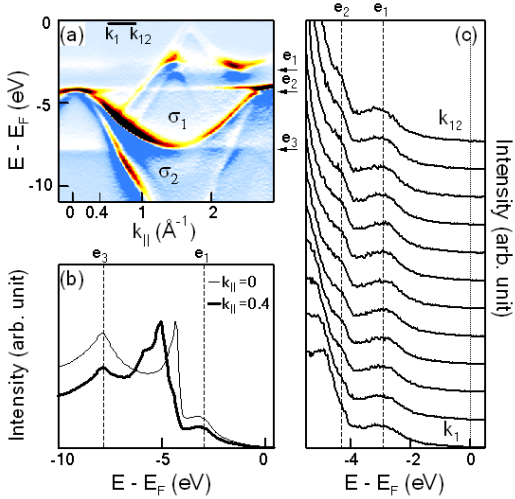}
\label{fig.11}
\caption{(a) First derivative of raw data taken at 55 eV photon energy (k$_z$ $\approx$ 0.11 c$^*$).  The color scale is the same as Figure 5.  Three nondispersive features are observed at energies e$_1$ (-2.9 eV), e$_2$ (-4.3 eV) and e$_3$ (-7.8 eV). (b) EDCs at $\Gamma$ point (thin line) and k$_\parallel$=0.4 $\AA^{-1}$ (thick line), showing the peaks at e$_1$ and e$_3$ are nondispersive.  (c) EDCs for the momentum range indicated by a horizontal line from k$_1$ to k$_{12}$, marked in panel a.} 
\end{figure}

In Figure 11, we introduce disorder/inhomogeneity features that are pronounced only in the case of HOPG samples.  Panel a shows the first derivative in energy of an ARPES map, from E$_F$ to -11 eV\@.  The first derivative is a method which allows to enhance the dispersive features as well as rising or falling edges in the data, and thus is particularly useful for detecting nondispersive peaks and edges.  In panel a we can clearly distinguish three nondispersive features at -2.9 eV, -4.3 eV and -7.8 eV, indicated by arrows on the same figure.  The nondispersive nature of these features can be checked by the EDCs shown in panels b and c.  These features, appearing as sharp extended horizontal lines in panel a suggests the presence of nondispersive localized states.  The energies of these nondispersive features occur at the top and bottom of the dispersive band and these features are strongly connected with the dispersive features associated with the $\pi$ and $\sigma$ bands.  Thus we have associated the presence of such nondispersive features  to the elastic scattering of electrons in either the initial state or the final state by inhomogeneity or disorder \cite{ZhouPRB}.

We now discuss the more important effect of disorder in the low energy electronic properties, as observed in both single crystal graphite samples and HOPG samples.  In order to investigate this effect, we have performed position dependent ARPES study with main focus on the near E$_F$ states.  Each spectrum is averaged over $\approx$ 100 $\mu m$, the spot size of the synchrotron beam.  We find that some of the low energy properties of the electronic structure are indeed strongly position dependent and we associate them with the presence of disordered states.  We note that similar behavior has been observed in different kinds of graphite samples measured, suggesting that these are important properties associated with graphite.

\begin{figure}
\includegraphics[width=8 cm]{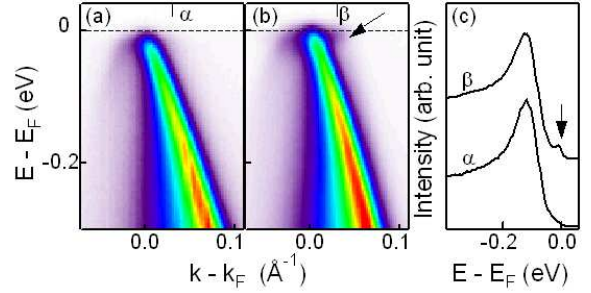}
\label{fig.12}
\caption{(a) ARPES intensity map taken at 26 eV photon energy (k$_z$ $\approx$ 0.13 c$^*$).  (b) ARPES intensity map measured at the same conditions as panel a except on a different spot inside the sample.  The arrow points to the additional intensity near E$_F$.  (c) EDCs at momentum values labeled as $\alpha$ and $\beta$ in panels a and b.  An additional peak is observed in EDC taken at $\beta$.} 
\end{figure}

In Figure 12, we report typical ARPES intensity maps near E$_F$, taken at 26 eV photon energy in different positions on the same sample.  While panel a show a parabolic $\pi$ band similar to the feature discussed in Figure 9, it is clear that, as we change position within the same sample, we observe an additional weakly-dispersive electron-like feature, within 50 meV below E$_F$ (panel b), coexisting with the parabolic $\pi$ band.  The presence of this additional electron-like feature in panel b can be clearly observed in the EDC shown in panel c.  Here an additional weak peak (pointed to by an arrow in panel c) is clearly observed.  Note that while the presence of the $\pi$ band is position independent, observed in all the samples studied for all positions measured, this additional electron-like feature is strongly position dependent, for each of the sample studied.

\begin{figure}
\includegraphics[width=9 cm]{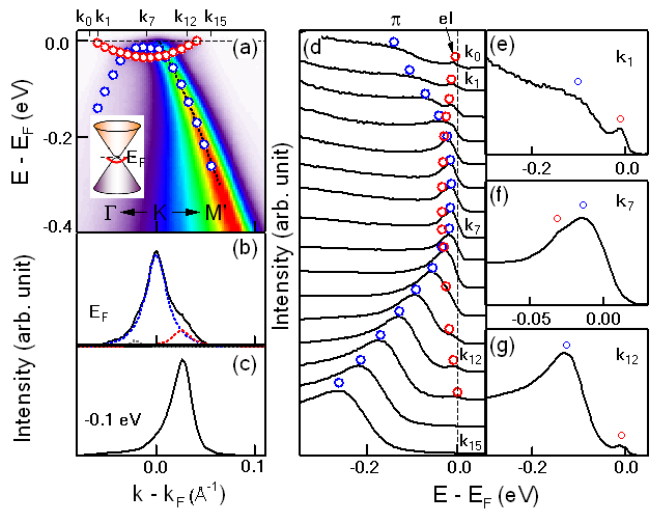}
\label{fig.13}
\caption{(a) ARPES intensity map taken at 26 eV photon energy (k$_z$ $\approx$ 0.13 c$^*$).  Red and blue circles are the dispersions extracted from EDCs for the electron-like band and $\pi$ bands, while the dotted black line is the dispersion extracted from MDCs. (b) MDC at E$_F$\@.  Blue, red and gray dotted lines are the peaks used to fit the MDC.  (c) MDC at -0.1 eV.  (d) EDCs taken at momenta from k$_0$ to k$_{15}$ indicated by vertical tick marks on the top of panel a.  The EDCs at k$_0$ to k$_5$ are scaled by different factors so that all the peaks can be seen.  (e-g) EDCs at k$_1$, k$_7$, k$_{12}$.}
\end{figure}

Figure 13 shows the detailed analysis of this additional band near E$_F$.  The details of this electron-like band and the $\pi$ band are analyzed from the MDCs and EDCs shown in panels b-g.  Panel b shows the MDC at E$_F$.  A main peak in the center, associated with the $\pi$ band, and two side peaks associated with the additional band, can be distinguished.  As the binding energy increases, the low energy band is no longer present and the peaks in the MDC at -100 meV (panel c) are from the main $\pi$ band.  However, the dispersions of the low energy bands are difficult to follow in the MDCs, partly due to the highly parabolic bands in this region that makes an MDC analysis hard to interpret.  On the other hand, in the EDCs, this low energy feature shows up clearly as a small well-defined peak at k$_1$ (panel e), k$_{12}$ (panel g) and  a small hump in k$_7$ (panel f) and thus we use EDC analysis (panel d) to extract the dispersion.  The extracted EDC dispersions are overplotted in panel a for both the low and high energy bands.  By fitting the electron pocket, we can directly measure the mass of the electrons.  This gives a value $\approx$ 0.4 m$_e$, which is much larger than that of electrons and holes as measured in transport \cite{ZhangPRL, SouleHall, GaltCR}.  Also, by estimating the volume of the large electron pocket \cite{fnlargeE}, we obtain an electron concentration of $\approx$ 8$\times$10$^{19}$ cm$^{-3}$.  This electron concentration is again an order of magnitude higher than the value reported by transport measurements \cite{ZhangPRL, SouleHall}.  

We now discuss the origin of this feature.  While it seems appealing to associate this electron-like feature with the electron pocket predicted by band structure at k$_z$=0, we note that its position dependence as well as the detailed analysis of the effective mass and carrier concentration, clearly suggests a different origin for this feature.  One possible explanation for this large electron pocket is due to defect-induced localized states \cite{Castroneto, Dresselhaus, LocState}, e.g.~states along a zigzag edge.  Additional support comes from STM, where a peak in the local density of states at an energy ($\approx$ -0.03 eV) similar to the weakly dispersing electron pocket discussed here \cite{STMedges}, is observed near zigzag edges.

\begin{figure}
\includegraphics[width=7.8 cm]{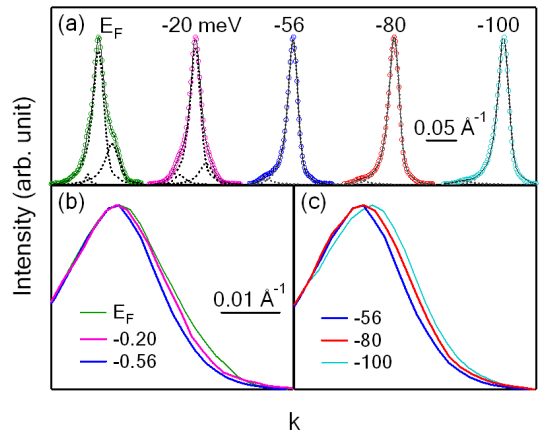}
\label{fig.14}
\caption{(a) Raw MDCs at energies from E$_F$ to -0.1 eV.  The dotted lines are the peaks used to fit the MDC.  The solid lines are the fit curves.  (c-d) Comparison of the central peak at different energies, between E$_F$ and -56 meV (panel c) and -56 meV and -100 meV (panel d), after subtraction of the side peaks.  The MDCs are shifted so that the half maximum positions on the left are at the same k point, which allows a direct comparison of full width half maximum by comparing the half maximum position on the right.  Interestingly the MDC peak near -56 meV has the smallest width.}
\end{figure}

To gain more insight into the nature of this low energy band, it is important to study MDC full width half maximum (FWHM), which gives information about the inverse lifetime or imaginary part of the self energy.  Figure 14(a) shows the quality of the MDC fit from which the FWHM is extracted.  Within the band width of the electron pocket (E$_F$ to -40 meV), the MDCs are fitted with 3 peaks.  For binding energy larger than the band width of the electron pocket (-44 meV to -100 meV), the MDCs are fitted with 2 peaks only.  Panels b and c show the comparison of the main peak from the $\pi$ band after subtracting the side peaks.  Surprisingly, the minimum MDC FWHM is at $\approx$ -56 meV (see comparisons in panels b, c), contrary to the Fermi liquid theory where the minimum FWHM is expected to be at E$_F$.  

Figure 15 shows the extracted MDC FWHM.  In panel b, from -100 meV to -56 meV, the MDC width decreases, while near E$_F$, a sudden increase of the MDC FWHM is observed.  This sudden increase is present near the energy where the large electron-pocket is observed and therefore most likely associated with it.  It has been predicted that in the presence of disorder-induced localized states, the imaginary part of the self energy shows a minimum at a finite binding energy rather than at E$_F$ (panel c) \cite{Castroneto}.  The similarity between panels b and c presents an intriguing possibility that the MDC width analysis is also indicative of the fact that the electron pocket stems from an impurity band induced by defects.  However, as we noted earlier, one of the caveats of the MDC analysis is the assumption of a linear band dispersion relation.  The breakdown of this assumption in the current case makes it necessary a more sophisticated line shape analysis, which is in progress.

\begin{figure}
\includegraphics[width=9 cm]{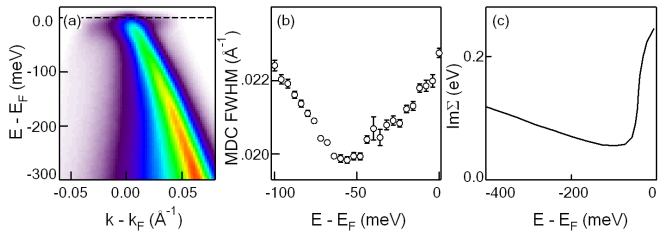}
\label{fig.15}
\caption{(a) ARPES intensity map taken at 26 eV.  (b) MDC full width half maximum at low energies between -100 meV and $E_F$.  An anomalous upturn around -56 meV can be observed. The error bar from the fit is included.  (c) Imaginary part of the electron self energy, including both the effect of electron-electron interaction and the effect of 0.01$\%$ disorder \cite{Castroneto}.}
\end{figure}

\section*{CONCLUSION}
In conclusion, we have presented a detailed high resolution ARPES study of the electronic band structure of graphite.  
We found that, while the overall band structure is in agreement with LDA calculation, the low energy excitations present new surprises. 
First of all, we report the first evidence that, as a result of the interlayer coupling a splitting of the $\pi$ bands occurs far from the Fermi level, with an overall splitting of 0.7 eV.  Second we observed that, as we move closer to the Fermi energy, the low energy excitations near the H point are Dirac-like massless fermions which coexist with quasiparticles with finite effective mass near the K point.  In addition, we report first ARPES signatures of electron-phonon interaction in graphite.  Finally, we also found that the electronic structure is strongly affected by lattice disorder resulting in strong defect-induced localized state near the Fermi energy, which strongly contribute to the quasiparticle scattering process.

\begin{acknowledgments}
We thank A. Castro Neto, Steven G. Louie, D. H. Lee and C. D. Spataru, for useful discussions.  This work was supported by the National Science Foundation through Grant No.~DMR03-49361 and by the Director, Office of Science, Office of Basic Energy Sciences, Division of Materials Sciences and Engineering of the U.S Department of Energy under Contract No.~DEAC03-76SF00098. 
\end{acknowledgments}

\end{document}